\documentclass[a4paper,12pt]{article}
\usepackage{epsfig}
\usepackage{amssymb}
\usepackage{amsfonts}
\usepackage{amsmath}
\usepackage{euscript}
\usepackage{verbatim}
\usepackage{latexsym}
\usepackage{graphicx}

\newskip\humongous \humongous=0pt plus 1000pt minus 1000pt

\newif\ifdtup

\catcode`\@=11

\@addtoreset{equation}{section}

\def\@normalsize{\@setsize\normalsize{15pt}\xiipt\@xiipt
\abovedisplayskip 14pt plus3pt minus3pt%
\belowdisplayskip \abovedisplayskip
\abovedisplayshortskip \z@ plus3pt%
\belowdisplayshortskip 7pt plus3.5pt minus0pt}

\def\small{\@setsize\small{13.6pt}\xipt\@xipt
\abovedisplayskip 13pt plus3pt minus3pt%
\belowdisplayskip \abovedisplayskip
\abovedisplayshortskip \z@ plus3pt%
\belowdisplayshortskip 7pt plus3.5pt minus0pt
\def\@listi{\parsep 4.5pt plus 2pt minus 1pt
     \itemsep \parsep
     \topsep 9pt plus 3pt minus 3pt}}

\relax

\catcode`@=12

\catcode`\@=11

\def\section{\@startsection{section}{1}{\z@}{3.5ex plus 1ex minus
   .2ex}{2.3ex plus .2ex}{\large\bf}}

\def\SymBoxes#1#2#3#4{\newdimen\un@t \un@t#3%
\raisebox{#1}{\rule{#2\un@t}{#4}\hskip-#2\un@t
\@tempdimb\un@t \advance\@tempdimb by-#4\@tempcntb#2\relax%
\@whilenum{\@tempcntb>0}\do{
\rule{#4}{\un@t}\hskip\@tempdimb \advance\@tempcntb by\m@ne}%
\hskip-#2\un@t \rule[\un@t]{#2\un@t}{#4}%
\rule[\un@t]{#4}{#4}\hskip-#4
\rule{#4}{\un@t}}\hskip-#4}                

\begin{document}

\newcommand{\beq}{\begin{equation}}
\newcommand{\eeq}{\end{equation}}
\newcommand{\bea}{\begin{eqnarray}}
\newcommand{\eea}{\end{eqnarray}}
\newcommand{\beas}{\begin{eqnarray*}}
\newcommand{\eeas}{\end{eqnarray*}}
\newcommand{\defi}{\stackrel{\rm def}{=}}
\newcommand{\non}{\nonumber}
\newcommand{\bquo}{\begin{quote}}
\newcommand{\enqu}{\end{quote}}
\renewcommand{\(}{\begin{equation}}
\renewcommand{\)}{\end{equation}}
\def \eqn#1#2{\begin{equation}#2\label{#1}\end{equation}}
\def\IZ{{\mathbb Z}}
\def\IR{{\mathbb R}}
\def\IC{{\mathbb C}}
\def\IQ{{\mathbb Q}}
\def\de{\partial}
\def\Tr{ \hbox{\rm Tr}}
\def\H{ \hbox{\rm H}}
\def\HE{ \hbox{$\rm H^{even}$}}
\def\HO{ \hbox{$\rm H^{odd}$}}
\def\K{ \hbox{\rm K}}
\def\Im{ \hbox{\rm Im}}
\def\Ker{ \hbox{\rm Ker}}
\def\const{\hbox {\rm const.}}
\def\o{\over}
\def\im{\hbox{\rm Im}}
\def\re{\hbox{\rm Re}}
\def\bra{\langle}\def\ket{\rangle}
\def\Arg{\hbox {\rm Arg}}
\def\Re{\hbox {\rm Re}}
\def\Im{\hbox {\rm Im}}
\def\exo{\hbox {\rm exp}}
\def\diag{\hbox{\rm diag}}
\def\longvert{{\rule[-2mm]{0.1mm}{7mm}}\,}
\def\a{\alpha}
\def\dag{{}^{\dagger}}
\def\tq{{\widetilde q}}
\def\p{{}^{\prime}}
\def\W{W}
\def\N{{\cal N}}
\def\hsp{,\hspace{.7cm}}

\def\br{\nonumber\\}
\def\IZ{{\mathbb Z}}
\def\IR{{\mathbb R}}
\def\IC{{\mathbb C}}
\def\IQ{{\mathbb Q}}
\def\IP{{\mathbb P}}
\def \eqn#1#2{\begin{equation}#2\label{#1}\end{equation}}

\newcommand{\C}{\ensuremath{\mathbb C}}
\newcommand{\Z}{\ensuremath{\mathbb Z}}
\newcommand{\R}{\ensuremath{\mathbb R}}
\newcommand{\rp}{\ensuremath{\mathbb {RP}}}
\newcommand{\cp}{\ensuremath{\mathbb {CP}}}
\newcommand{\vac}{\ensuremath{|0\rangle}}
\newcommand{\vact}{\ensuremath{|00\rangle}                    }
\newcommand{\oc}{\ensuremath{\overline{c}}}
\begin{titlepage}
\begin{flushright}
CHEP XXXXX
\end{flushright}
\bigskip
\def\thefootnote{\fnsymbol{footnote}}

\begin{center}
{\Large
{\bf Attraction, with Boundaries 
}
}
\end{center}

\bigskip
\begin{center}
{\large  Avik Chakraborty$^a$\footnote{\texttt{avikchakraborty88@yahoo.in}} and
Chethan Krishnan$^a$\footnote{{\texttt{chethan@cts.iisc.ernet.in}}}}
\vspace{0.1in}

\end{center}

\renewcommand{\thefootnote}{\arabic{footnote}}

\begin{center}
$^a$ {Center for High Energy Physics\\
Indian Institute of Science, Bangalore, India}\\

\end{center}

\noindent
\begin{center} {\bf Abstract} \end{center}
We study the basin of attraction of static extremal black holes, in the concrete setting of the STU model. By finding a connection to a decoupled Toda-like system and solving it exactly, we find a simple way to characterize the attraction basin via competing behaviors of certain parameters. The boundaries of attraction arise in the various limits where these parameters degenerate to zero. We find that these boundaries are generalizations of the recently introduced (extremal) subtracted geometry: the warp factors still exhibit asymptotic integer power law behaviors, but the powers can be different from one. As we cross over one of these boundaries (``generalized subttractors"), the solutions turn unstable and start blowing up at finite radius and lose their asymptotic region. 
Our results are fully analytic, but we also solve a simpler theory where the attraction basin is lower dimensional and easy to visualize, and present a simple picture that illustrates many of the basic ideas. 

\vspace{1.6 cm}
\vfill

\end{titlepage}

\setcounter{footnote}{0}

\section{Introduction}
\label{intro}

In this paper, we will consider attractor black holes \cite{Goldstein, FKS, SenReview, DabholkarSen}: specifically, we will study the basins of attraction of static non-supersymmetric attractors mostly in STU supergravity. Attractor black holes are extremal black holes that can exhibit scalar hair for a fixed horizon value of the charges. Since the thermodynamic properties of the black hole are determined by its charges, the attraction phenomeon is a hint that only the near-horizon geometry of the black hole might matter for understanding its thermodynamics.

In a recent paper \cite{Avik}, we showed that the recently introduced subtracted geometry \cite{CL1, CL2, CG} (in its extremal limit) can be thought of as a boundary of an attraction basin. Subtracted geometries are black holes in Einstein-Maxwell-dilaton theories that arise upon replacing a certain warp factor in the metrics of flat space black hole solutions by another (specific) warp factor. The precise algorithm for the replacement depends on the original black hole metric, and the replacement changes the asymptotics. But it does not change the black hole's thermodynamics and it manifests a certain ``hidden" conformal symmetry of the original black hole. The original motivations for introducing the subtracted geometry can be found eg., in \cite{CveticLarsenOld, CKreview, StromKerrCFT, LMP, Hartman, Stanislav, Geoffrey, HiddenCS, CK}, but we will not concern ourselves with that here. Our goal will be to study the attraction basins of some exactly solvable attractor systems analytically, and to identify the various boundary behaviors that emerge. We find that the extremal subtracted (``subttractor") geometry that was found in \cite{Avik} and some generalizations of it arise naturally as boundaries of the attraction basin: the original subtracted geometry of \cite{CL1, CL2, CG}  had a warp factor that went linearly in $r$ for large $r$, but we find that in the STU model, attractor boundaries can have behaviors that go as $\sim r, r^2$ or $r^3$. We call them generalized subttractors. They can all be understood in terms of replacements of the warp factor in the original geometry. 

The systems we consider are fully integrable in terms of a Toda system, so we can find exact solutions and get a complete understanding of the attraction basin. 
We can characterize the general solutions in terms of certain integration constants $d_a$. The solutions are regular everywhere at and outside the horizon only when $d_a \geq 0$. Therefore, the entire attraction basin is captured in $d_a$-space by the first quadrant for a simple one-dilaton theory where $d_a$ can be $d_1$ or $d_2$ and by the first orthant in the STU model where $a$ can take four values. It will be interesting to see what generalizations of these statements can be made for other Einstein-Maxwell-dilaton theories, but the question is not necessarily straightforward: we got lucky here because the system is Toda integrable. 

We find that the boundaries of the attraction basin occur when one (or more) of the $d_a$'s degenerate to zero. We provide a full characterization of such boundaries. 
The unstable solutions corresponding to negative $d_a$ do not have an asymptotic region and diverge at finite radius. 

The plan of this paper is as follows. In the next section, we describe our system: the black holes we consider all fall into a certain attractor ansatz (even though all the black holes allowed by the ansatz are not extremal/attractive). We show that under certain circumstances, the coupled system of Einstein-Maxwell-dilaton equations of motion can be brought to the form of a generalized Toda system and can be integrated exactly. The systems we consider fall into this category, and can be diagonalized and integrated. In section \ref{STUtoda}, we introduce our main object of interest, namely static black holes in a consistent truncation of STU supergravity. We solve the system completely\footnote{We emphasize that the solution has previously been found in a different form in \cite{JdBoer}, but we will describe some advantages of our approach when the context arises.} using our Toda approach and describe the attraction basin and the boundaries in detail. Since the STU system has three scalars and the attraction basin in terms of the $d_a$ parameters is four dimensional, we then turn to a simpler system. This system has the advantage that some basic ideas remain the same, yet we can get a more intuitive feel for the attraction basin. The theory we consider can be obtained from the STU system \cite{Avik} and is also integrable and we find a fully consistent picture. 

The solution space is quite rich and we find various auxiliary results on the way but we have relegated them to appendices. Among these are the observation that the ``vertex of attraction" is an $AdS_2 \times S^2$ space, and the identification of the slice along which a given black hole flows in the attraction basin when perturbed. 
We conclude with some comments about various connections and open questions.

\section{A General Toda-like System}
\label{Toda}

A general Lagrangian arising in (ungauged) supergravity is
\bea \label{action}
S=\int d^4x \sqrt{-g}\Big(R-2 (\partial \phi_i)^2- f_{ab}(\phi_i)F^{a}_{\mu\nu}F^{b \ \mu\nu} -\frac{1}{2}\tilde f_{ab}(\phi_i))F^{a}_{\mu\nu}F^{b}_{\rho \sigma}\epsilon^{\mu\nu\rho\sigma}\Big)
\eea
where $i=1,...,n$ is the number of scalars and $a,b=1,...,N$ counts the number of $U(1)$ gauge fields. The $f_{ab}$ and $\tilde f_{ab}$ are the gauge couplings including the axionic piece. This is the form considered in \cite{Goldstein}. We will look for solutions of this system in the ``attractor ansatz", which means we will consider solutions of the form
\bea
ds^2=-a(r)^2 dt^2+\frac{dr^2}{a(r)^2}+b(r)^2 d \Omega^2, \hspace{1in} \\
F^a=Q^a_m\sin \theta d\theta \wedge d\phi+ \frac{ f^{ab}(\phi_i)}{b^2}(Q_{eb}-\tilde f_{bc}Q^c_m)dt \wedge dr,\ \ \ \ \phi_i \equiv \phi_i(r). \label{gauge-scalar}\hspace{ 0.1in}
\eea
The equations of motion take the form
\bea\label{minansatze1}
(a^2\ b^2)''-2 =0 \label{att1} \\
\frac{b''}{b}+{\phi_i'}^2=0 \\
(a^2 b^2\phi_i')'-\frac{\partial_{\phi_i} V_{\rm eff}(\phi_i)}{2 b^2}=0
\eea
and the first order constraint from Einstein equations is
\bea 
a^2{b'}^2+{a^2}'{b^2}'+\frac{V_{\rm eff}(\phi_i)}{b^2}-a^2b^2 {\phi_i'}^2-1=0. \label{minansatze2}
\eea
The effective potential that shows up in the equations of motion is
\bea\label{mineff}
V_{\rm eff}(\phi_i)=f^{ab}(\phi_i)(Q_{ea}-\tilde f_{ac}(\phi_i)Q^c_m)(Q_{eb}-\tilde f_{bd}(\phi_i)Q^d_m)+f_{ab}(\phi_i)Q^a_mQ^b_m. \label{effpot}
\eea

At this stage, generalizing \cite{Goldstein} we introduce
\bea
u_i=\phi_i \ \ {\rm for}\ \ i=1,..,n, \ \ \ u_{n+1}=\log a, \ \ z=\log ab.
\eea
With these definitions, via simple linear combinations, one can bring the equations to the form
\bea
\ddot u_{n+1}&=&e^{2 u_{n+1}}V_{\rm eff}(u_i), \label{main1}\\
\ddot u_i&=& \frac{1}{2}e^{2 u_{n+1}}\partial_{u_i}V_{\rm eff}(u_i), \label{main2}\\
\ddot z &=& e^{2 z}.
\eea
The energy constraint is redundant and takes the form
\bea
\dot u_i^2+\dot u_{n+1}^2+e^{2 z}-\dot z^2 =e^{2 u_{n+1}}V_{\rm eff} (u_i).
\eea
The dots stand for derivatives with respect to the tortoise coordinate
\bea
\tau=\int \frac{dr}{a^2 \ b^2}=\frac{1}{2m}\log\Big(1-\frac{2m}{r}\Big) \label{tortoise}
\eea
We choose boundary conditions such that the solutions are regular at $r=2m$. This is what we call the outer horizon and we choose our notations so that they match with the black holes in \cite{CG, Chong} in the appropriate limits. The inner horizon has been fixed by a coordinate choice to be at $r=0$. The energy constraint has been used in bringing the first three equations to the simple form presented here. The equation for $z$ is really the statement that $(a^2 b^2)''=2$ and is solved by 
\bea
b^2=r(r-2m)/a^2,
\eea
once we make our coordinate choices. So (\ref{main1}-\ref{main2}) are what we really need to solve.

So far our set up is completely general within the attractor ansatz. To proceed further, we will assume a less general form for $V_{\rm eff}$. The form we will consider will still be general enough to contain the previous exact solutions \cite{Goldstein, Avik} and as we will show in the next section, the STU attractor also falls in this class:
\bea
V_{\rm eff}(u_i)=\sum_{a=1}^N \Big(Q_a^2 e^{\sum_{i=1}^n\alpha_{a i} u_i}\Big).
\eea
As explicitly indicated, there is one summation over $a$ and another one over $i$. Now, we will assume that 
\bea
N=n+1.\label{Nn}
\eea
The notation in the following will be that summation over repeated indices (irrespective of how many times they are repeated or the precise index placement) will run from $1, ..., N$ for $a, b, c,d$ and from  $1, ..., n$ for $i, j, k$.   Using this, and defining\footnote{We can (and will) think of $\alpha_{ab}$ as a square matrix $\alpha$ or as a set of row vectors $\alpha_a$ depending on the context. In (\ref{2a}), $2_a$ stands for a row vector indexed by $a$ in which each slot contains the number 2.} 
\bea
\alpha_{ab}\equiv (\alpha_a)_b\equiv (\alpha_{ai} \ \ \alpha_{aN})=(\alpha_{ai} \ \ 2^T_a) \label{2a}
\eea
brings (\ref{main1}-\ref{main2}) to the form
\bea 
\ddot u_i&=&\frac{1}{2} Q_a^2\ \alpha_{a i}\ e^{\alpha_{ab} u_b} \\
\ddot u_N&=& Q_a^2 \ e^{\alpha_{ab} u_b}. 
\eea
With our choice of the effective potential and because of the relation (\ref{Nn}), these two equations can be combined to the form 
\bea
\ddot u_a=A_{a c}\ Q_c^2\ e^{\alpha_{cb} u_b}
\eea
where the square matrix $A$ is defined via
\bea
A=\left(
\begin{array}{c}
\frac{1}{2}\alpha^T_{ai} \\
1_a
\end{array}
\right),
\eea
and $a$ is the number of the column. Defining the matrix $m_{ab}=\alpha_{ac}  A_{cb}$ and making the change of variables 
\bea
X_a=(A^{-1})_{ab}u_b+(m^{-1})_{ab} \ln Q^2_b
\eea
brings us to the final form
\bea
\ddot X_a=e^{m_{ab}X_b}.
\eea
This is a set of $N$ differential equations which take a simple structure: they are coupled to each other linearly in the exponent. The matrix $m_{ab}$ is a constant matrix, and therefore our equations can typically be related to Toda systems via simple tricks (see eg. \cite{Goldstein, Avik, GibMaed, Pope}). We will do this for static solutions in the STU model in the next section.

\section{Toda Solutions in the STU Model}
\label{STUtoda}

We will mostly be working with a theory that has three scalars and four vectors in most of this paper, that arises as a consistent truncation of ${\cal N}=2$ supergravity with four vector multiplets \cite{Avik, CG, Chong}\footnote{We can also think of the system as arising in the STU model, up to an electric-magnetic duality. At the level of the effective attractor ansatz theory that we write down, this makes no difference.}:   
\bea
\int d^4x \sqrt{-g}\Big[R- \frac{1}{2} (\partial \varphi_i)^2-\frac{1}{4}\Big(e^{-\varphi_1+\varphi_2-\varphi_3}( F^{1}_{\mu\nu})^2+e^{-\varphi_1+\varphi_2+\varphi_3}(F^{2}_{\mu\nu})^2+\nonumber \hspace{0.5in}\\ 
\hspace{0.5in}+e^{-\varphi_1-\varphi_2+\varphi_3}({\cal F}^{1}_{\mu\nu})^2+e^{-\varphi_1-\varphi_2-\varphi_3}({\cal F}^{2}_{\mu\nu})^2\Big)\Big]
\eea 
We look for an attractor ansatz for this action of the form presented in the previous section, with the specific choice for
the gauge fields:
\bea
F_1=2 Q_1 \sin \theta d\theta \wedge d\phi, \ && \ F_2=4 e^{2(\phi_1-\phi_2-\phi_3)}\frac{Q_2}{2b(r)^2}dt \wedge dr, \\
F_3\equiv{\cal F}^{1}=2 Q_3 \sin \theta d\theta \wedge d\phi, \ && \ F_4\equiv{\cal F}^2=4 e^{2(\phi_1+\phi_2+\phi_3)}\frac{Q_4}{2b(r)^2}dt \wedge dr,
\eea
This is of the ansatz presented in (\ref{gauge-scalar}).  
With these, the equations of motion take the form presented in the previous section, with $\varphi_i=2 \phi_i$, and with the effective 
potential
\bea
V_{\rm eff}=Q_1^2e^{-2 \phi_1+2 \phi_2-2 \phi_3}+Q_2^2e^{2\phi_1-2 \phi_2-2 \phi_3}+Q_3^2e^{-2\phi_1-2\phi_2+2\phi_3}+Q_4^2e^{2\phi_1+2 \phi_2+2\phi_3} \label{STUVeff}
\eea
This structure falls in the general form for the effective potential that we presented in the previous section. In fact with the choice 
\bea
u_1=\phi_1, \ \ u_2=\phi_2, \ \ u_3=\phi_3, \ \ u_4=\log a
\eea
the matrix 
\bea
A=\left(
\begin{array}{cccc}
-1 &+1&-1&+1\\
+1 &-1&-1&+1\\
-1 &-1&+1&+1\\
+1 &+1&+1&+1
\end{array}
\right),
\eea
and the $\alpha_{ab}$ matrix can be read off from it. This results in the $m_{ab}$ matrices becoming diagonal, and therefore the system decoupling into four separate ODEs:
\bea
\ddot X_a=e^{2 X_a}.
\eea
This is a trivial version of the Toda lattice equations where there is no coupling between neighbors and can be immediately solved by noting that $\ddot X=v \frac{dv}{dX}$, where $v\equiv \dot X$. The end result is
\bea
X_a=\log\Big(\frac{c_a}{\sinh c_a(\tau-d_a)}\Big)\equiv \log \Big(\frac{c_a}{F_a}\Big).
\eea
where $c_a$ and $d_a$ are integration constants (no summation over repeated indices here) and we have defined the $F_a$ in terms of the $\sinh$ functions for future convenience. This is the most general solution of the attractor ansatz in the STU model with effective potential given by (\ref{STUVeff}).

The above decoupling was observed in a different, but equivalent set up in the STU model in \cite{JdBoer}. It seems they came to this result by inspection and cleverness, we see here that it can be accomplished by cranking our Toda-ization machinery. The reason why we prefer our way here is that the integration constants that naturally arise in our set up have an immediate and clean interpretation in terms of the attraction basin. Of course, even though the forms of our solutions look superficially different, we show that the \cite{JdBoer} solution can be brought to our form by connecting our parameters to theirs carefully  (\ref{jdbconnection1}-\ref{jdbconnection2}).

Translating back from the Toda-style variable $X_a$ to the original variables, we find that 
\bea
e^{4 \phi_1}=\frac{Q_1\ Q_3}{Q_2 \ Q_4}\ \frac{c_2 \ c_4}{c_1\ c_3}\  \frac{F_1\ F_3}{F_2 \ F_4}, \hspace{0.8in}\\
e^{4 \phi_2}=\frac{Q_2\ Q_3}{Q_1 \ Q_4}\ \frac{c_1 \ c_4}{c_2\ c_3}\  \frac{F_2\ F_3}{F_1 \ F_4}, \hspace{0.8in}\\
e^{4 \phi_3}=\frac{Q_1\ Q_2}{Q_3 \ Q_4}\ \frac{c_3 \ c_4}{c_1\ c_2}\  \frac{F_1\ F_2}{F_3 \ F_4}, \hspace{0.8in}\\
a^2=\sqrt{\frac{c_1\ c_2\ c_3 \ c_4}{2^4\ Q_1\ Q_2 \ Q_3 \ Q_4}\frac{1}{F_1\ F_2\ F_3 \ F_4}}, \ \ b^2=\frac{r(r-2m)}{a^2}.
\eea

These solutions remain invariant under the $c_a \leftrightarrow -c_a$, so we will work with $c_a >0$. We want regularity as $r \rightarrow 2m$ (or, $\tau \rightarrow -\infty$). The finiteness of the scalars yield
\bea
c_1+c_3-c_2-c_4=c_2+c_3-c_1-c_4=c_1+c_2-c_3-c_4=0
\eea
which forces
\bea
c_1=c_2=c_3=c_4 \equiv c. 
\eea
The finiteness of $b$ at the horizon (which is necessary because otherwise the horizon will be singular) then gives us
\bea
c=m.
\eea
So the solutions which are regular at the horizon are
\bea
e^{4 \phi_1}=\frac{Q_1\ Q_3}{Q_2 \ Q_4}\   \frac{\sinh m(\tau-d_1)\ \sinh m(\tau-d_3)}{\sinh m(\tau-d_2) \ \sinh m(\tau-d_4)}, \hspace{0.8in}\\
e^{4 \phi_2}=\frac{Q_2\ Q_3}{Q_1 \ Q_4}\ \frac{\sinh m(\tau-d_2)\ \sinh m(\tau-d_3)}{\sinh m(\tau-d_1) \ \sinh m(\tau-d_4)}, \hspace{0.8in}\\
e^{4 \phi_3}=\frac{Q_1\ Q_2}{Q_3 \ Q_4}\ \frac{\sinh m(\tau-d_1)\ \sinh m(\tau-d_2)}{\sinh m(\tau-d_3) \ \sinh m(\tau-d_4)}, \hspace{0.8in}\\
a^2=\frac{m^2}{2^2}\frac{(Q_1Q_2Q_3Q_4)^{-1/2}}{\sqrt{\sinh m(\tau-d_1)\ \sinh m(\tau-d_2)\ \sinh m(\tau-d_3) \ \sinh m(\tau-d_4)}}, 
\eea
These solutions are general. In particular, we haven't imposed asymptotic flatness or extremality at the horizon. We can show that when all the $d_a$'s are positive, the solutions are regular everywhere outside the event horizon (which also is regular), and  asymptotically flat. But first we compare this to the solutions of \cite{JdBoer}. Remembering the relation (\ref{tortoise}) between $\tau$ and $r$, it can be shown that our solutions turn into theirs (see equations (2.32-2.35) in \cite{JdBoer}) when we set our scalars $\phi_i\equiv \eta_i/2$ (where $\eta_i$ are their scalars) and choose
\bea
Q_1=B_3/2, \ Q_2=B_1/2, \ Q_3=B_2/2, Q_4=q_0/2.\hspace{0.25in} \label{jdbconnection1}\\
\frac{2 \sinh md_1}{\cosh md_1-\sinh md_1}=a_3^2, \
\frac{2 \sinh md_2}{\cosh md_2-\sinh md_2}=a_1^2, \\
\frac{2 \sinh md_3}{\cosh md_3-\sinh md_3}=a_2^2, \ 
\frac{2 \sinh md_4}{\cosh md_4-\sinh md_4}=a_0^2 \label{jdbconnection2}
\eea
where the right hand sides are in the notations of \cite{JdBoer} and the left hand sides are our notations. The superficially strange assignment in the index matching arises because the theory has a cyclic symmetry in the $a=1,2,3 \ {\rm (but \ not \ }4{\rm )}$ index and their choice is related to ours up to a cyclic shift. Another comment is that their gauge fields are different from ours in the choice of the electric-magnetic duality frame, but the effective attractor theory is the same either way. 

For our purposes, our form of the integration constants is much more convenient. This is because as we will show presently, they are directly related to the data at asymptotic infinity. The \cite{JdBoer} integration constants involve $m$ which is an IR (near-horizon) piece of data. This mixed UV-IR notation for the data makes things less convenient to understand and work with\footnote{in particular, note that the extremal limit involves $m\rightarrow 0$ and the coefficients $a_0, ... a_4$ in \cite{JdBoer} collapse to zero in that limit. This indicates that these are not a good notation for working with attractors.}, so we will stick to our Toda-notation. 

Now we will demand that the geometry is asymptotically flat and that the scalars take finite values at infinity, ie., we demand 
\bea
a \stackrel{r \rightarrow \infty}{\longrightarrow} a_0, \ \ \phi_i  \stackrel{r \rightarrow \infty}{\longrightarrow}\phi_{i\infty}
\eea
We can use these to fix $d_a$ easily enough in the non-extremal case, but since we are interested in attractors, we will do this in the limit $m\rightarrow 0$. We get three equations from the scalars and one from $a (r)$, so we can fix the four integration constants via
\bea
d_1=\frac{1}{2\ a_0\ |Q_1|}\ e^{+\phi_{1\infty}-\phi_{2\infty}+\phi_{3\infty}},\label{asymps1}\\
d_2=\frac{1}{2\ a_0\ |Q_2|}\ e^{-\phi_{1\infty}+\phi_{2\infty}+\phi_{3\infty}},\\
d_3=\frac{1}{2\ a_0\ |Q_3|}\ e^{+\phi_{1\infty}+\phi_{2\infty}-\phi_{3\infty}},\\
d_4=\frac{1}{2\ a_0\ |Q_4|}\ e^{-\phi_{1\infty}-\phi_{2\infty}-\phi_{3\infty}},\label{asymps2}
\eea
The absolute value signs are there to ensure that if any of the $Q_a$'s are negative\footnote{Note that the possibilities for the signs of $Q_a$'s are such that one can have all of them positive, all of them negative or any two of them negative while the other two positive. Other choices are not allowed.} the corresponding $d_a$ will have an overall negative sign in its expression. This is necessary to make sure that all the $d_a$'s are positive, which is a necessary condition we will further elaborate on in the next section.

We have left the boundary value of $a_0$ to be an arbitrary (positive) constant at this stage, but in the attractor system every term that contains $a$ also contains a derivative of $r$, so the asymptotic value of the metric can be rescaled to unity by absorbing it in $r$. So these are asymptotically flat solutions. The fact that the asymptotic value of the scalar can be arbitrary is a direct consequence of the attraction phenomenon. Now we turn to a characterization of the attraction basin.

\section{Basin of Attraction}
\label{basins}

Even though the asymptotic values of the scalar can be arbitrary, the scalar perturbations at the horizon cannot be arbitrarily large. When they become too large, the solution stops flowing to asymptotically flat solutions, see Appendix \ref{subtslice}. Indeed, we will show that such solutions do not have an asymptotic region. This is what characterizes the boundaries of the attraction basin. We will try to understand them in this section.

In the extremal limit ($m\rightarrow 0$), our solutions take the form
\bea
e^{4 \phi_1}=\frac{Q_1\ Q_3}{Q_2 \ Q_4}\   \frac{(1+d_1 r)\ (1+d_3 r)}{(1+d_2 r) (1+d_4 r)}, \hspace{0.5in}\label{extremalexact1}\\
e^{4 \phi_2}=\frac{Q_2\ Q_3}{Q_1 \ Q_4}\ \frac{(1+d_2r)\ (1+d_3 r)}{(1+d_1 r) \ (1+d_4 r)}, \hspace{0.5in}\\
e^{4 \phi_3}=\frac{Q_1\ Q_2}{Q_3 \ Q_4}\ \frac{(1+d_1 r)\ (1+d_2 r)}{(1+d_3 r) \ (1+d_4 r)}, \hspace{0.5in}\\
a^2=\frac{1}{4}\frac{(Q_1Q_2Q_3Q_4)^{-1/2}\ r^2}{\sqrt{(1+d_1 r)\ (1+d_2 r)\ (1+d_3 r) \ (1+d_4 r)}}.\label{extremalexact2}
\eea
A crucial observation is that if we demand that $a$ is regular for all values of $r \in (0,\infty)$, then we are forced to choose all $d_a \ge 0$.  For negative values of any of the $d_a$, there will be a divergence at $r=-1/d_a >0$.

The sign choices of $Q_a$'s result merely in some trivial changes in the solution, so we will assume in what follows that all the $Q_a$'s are positive. In particular, note that the non-negativity of the $d_a$'s is independent of the signs of the $Q_a$'s.

We claim that the boundaries of the attraction basin correspond to some of the $d_a$'s degenerating to zero. It is clear from the forms of the scalars in  (\ref{asymps1}-\ref{asymps2}), or from taking the $r \rightarrow \infty$ limit of the (\ref{extremalexact1}-\ref{extremalexact2}) equations, that these correspond to the boundary values of the scalars diverging\footnote{The generalized subttractor boundaries arise as logarithmic envelope curves along which the scalar diverges, so there is no contradiction with the fact that the attraction basin has boundaries.}.

We classify the various possibilities below.
\begin{itemize}
\item[$\diamondsuit$] $d_4 \neq 0$ case.
\begin{itemize}
\item[$\bullet$] $d_1=d_2=d_3=0$ corresponds to $\phi_{1 \infty}, \phi_{2 \infty}, \phi_{3 \infty} \rightarrow -\infty$.
\item[$\bullet$] $d_1=d_2=0$, $d_3 \neq 0$ corresponds to $\phi_{1 \infty}, \phi_{2 \infty}$ fixed (or diverging slower than $|\phi_{3 \infty}|$), with $\phi_{3 \infty} \rightarrow -\infty$ faster than the others. The other cases where two of the $\phi_{i \infty} (i=1,2,3)$ are zero while the third one is not, can be obtained by symmetry.
\item[$\bullet$] $d_1=0$, $d_2 \neq 0 \neq d_3$ corresponds to $\phi_{1 \infty}, \phi_{3 \infty} \rightarrow -\infty$, with $\phi_{2 \infty} \rightarrow +\infty$. The other cases can be obtained by symmetry.
\item[$\bullet$] None of $d_1, d_2, d_3$ are zero. It is not a boundary.
\end{itemize}
\item[$\diamondsuit$] $d_4 = 0$ case.
\begin{itemize}
\item[$\bullet$] $d_1=d_2=d_3=0$. This solution corresponds to the ``vertex" of the attraction basin, and in fact is an $AdS_2 \times S^2$ geometry. We discuss it in an Appendix.
\item[$\bullet$] $d_1=d_2=0$, $d_3 \neq 0$ corresponds to $\phi_{1 \infty}, \phi_{2 \infty} \rightarrow \infty$, with $\phi_{3 \infty} \rightarrow -\infty$. The other cases can be obtained by symmetry.
\item[$\bullet$] $d_1=0$, $d_2 \neq 0 \neq d_3$ corresponds to $\phi_{1 \infty}, \phi_{3 \infty}$ fixed (or diverging slower than $|\phi_{2 \infty}|$), with $\phi_{2 \infty} \rightarrow +\infty$ faster than the others. The other cases are obtained by symmetry.
\item[$\bullet$] None of $d_1, d_2, d_3$ are zero, corresponds to $\phi_{1 \infty}, \phi_{2 \infty}, \phi_{3 \infty} \rightarrow +\infty$.
\end{itemize}
\end{itemize}

The restriction of positivity on $d_a$'s implies that asymptotically flat (attractor) solutions fall in the first orthant of the $(d_1, d_2, d_3, d_4)$ space. The boundaries of the orthant correspond to the attraction boundaries. When one of the $d_a$'s becomes less than zero, we end up with solutions that diverge at finite radius. We will illustrate these matters using a simpler system in the next section. 

We present in an appendix the explicit forms of these various attraction boundaries. We find that they can all be understood as new solutions with new warp factors (\` a la ``subtracted geometry" \cite{CL1, CL2, CG}) on the original Cvetic-Youm black hole, but with more general replacements of the warp factor than the original subtracted geometry. The asymptotic behavior of the warp factor can take different powers of $r$ : it can go as $r, r^2$ or $r^3$. The case when it goes as $r$ corresponds to the original subtracted geometry of \cite{CL2}. Note that the structure of our extremal solution (\ref{extremalexact1}-\ref{extremalexact2}) allows also the asymptotic behavior $r^0$ and $r^4$. The former corresponds to a unique $AdS_2 \times S^2$ solution (which we discuss in Appendix \ref{ads}). The latter represents the usual flat space hairy (and therefore highly degenerate) solutions. The in-between powers $r, r^2$ or $r^3$ correspond to various attraction boundaries with varying degrees of degeneracy corresponding to edges (of various dimensions) of the basin. We discuss these general subttractors in Appendix \ref{gsubtt}.

We note that the standard Cvetic-Youm black hole's \cite{CveticYoum, CveticYoum2, Chong} extremal limit (see Appendix where we quote this solution in its extremal limit) corresponds to choosing $ d_a=\frac{1}{2 |Q_a|}$.
Another interesting case is when $d_1=d_2=d_3=d_4$. This is the hairless extremal black hole (which is different from the extremal Cvetic-Youm solution). These are black holes in the central positively directed ray in the attraction basin spanned by $d_1,..., d_4$.


The structure of the attraction basin in the STU model is pretty intricate because of the number of scalars $\phi_i$ ($=3$) and the number of parameters $d_a$ ($=$ number of charges $=4$). In the next section we will investigate a simpler system with only one scalar and two charges where the attraction basin is much easier to visualize. We will see that the general picture that emerges is very similar.

\section{Toy Model}
\label{toy}

We will work with the model considered in \cite{Avik}. The action is of the form (\ref{action}), with 
\bea
f_{ab}(\phi)=\left(
\begin{array}{cc}
e^{\alpha_1 \phi}& 0 \\
0& e^{\alpha_2 \phi}
\end{array}
\right),
\eea
and $(\alpha_1, \alpha_2)=(2\sqrt{3}, -2/\sqrt{3})$. 
The axionic coupling matrix $\tilde{f}_{ab}$ is zero. The attractor ansatz and the equations of motion goes through as before with the understanding that the charges are two in number and both magnetic, and the index $i$ corresponds to a single scalar. The effective potential in this case also has a structure that enables Toda-ization as well as diagonalization \cite{Goldstein}\footnote{This can be understood because this system has a connection with the STU model \cite{Avik} which we showed to be diagonalizable after Toda-ization in this paper.}. This means that we can find exact solutions here as well, and the general solutions which are regular at an extremal horizon can be obtained by taking the $m \rightarrow 0$ limits of equations (4.1-4.3) in \cite{Avik}. The result is
\bea
e^{4 \phi/\sqrt{3}}=\frac{1}{\sqrt{3}}\frac{Q_2\ (1+d_2 r)}{Q_1 (1+d_1 r)},\hspace{0.83in} \label{singlescalar} \\
a^2 =\frac{r^2}{\Xi\sqrt{Q_1 Q_2^3}}\frac{1}{\sqrt{(1+d_1 r)(1+d_2 r)^3}}, \ \ b^2=r^2/a^2,\label{singlescalar2}
\eea
where
\bea
\Xi=\Big(-\frac{\alpha_2}{\alpha_1}\Big)^\frac{\alpha_1}{\alpha_1 - \alpha_2} + \Big(-\frac{\alpha_1}{\alpha_2}\Big)^\frac{-\alpha_2}{\alpha_1 - \alpha_2},
\eea
is a purely numerical factor. 
The structure here is obviously very similar to the one we found in the last section in the STU model, but here there are only two parameters $d_a=d_1, d_2$ and only one scalar. As before, the regular asymptotically flat solutions exist only in the $d_a>0$ regime. Unlike the four dimensional space of STU model however, here the $d_a$'s span a two-dimensional space and the attraction basin is the 1st quadrant. So it is easy to plot useful pictures of the basin here. 

The boundaries are at $d_1=0$ and $d_2=0$. By setting $a \stackrel{r \rightarrow \infty}{\longrightarrow} a_0, \ \ \phi_i  \stackrel{r \rightarrow \infty}{\longrightarrow}\phi_{i\infty}$, we find
\bea
d_1=\frac{3^{-3/8}}{a_0\sqrt{\Xi} |Q_1|}e^{{-\phi_\infty}\sqrt{3}}, \ \ 
d_2=\frac{3^{1/8}}{a_0\sqrt{\Xi} |Q_2|}e^{{\phi_\infty}/\sqrt{3}},
\eea
which implies that $d_1=0$ corresponds to the upper boundary where $\phi_\infty \rightarrow +\infty$ and $d_2=0$ corresponds to the lower boundary $\phi_\infty \rightarrow -\infty$. It is easy to check from (\ref{singlescalar}) that this latter solution corresponds to conventional subtracted geometry in the extremal limit (``subttractor"), and it was arrived at in this way in \cite{Avik}. The warp factor there goes as $\sim r$. The $d_1=0$ boundary however corresponds to $\phi_\infty \rightarrow +\infty$ and it is a new kind of subtracted geometry and the warp factor goes as $\sim r^3$ in the asymptotic region. 

\begin{figure}
\begin{center}
\includegraphics[height=0.35\textheight]{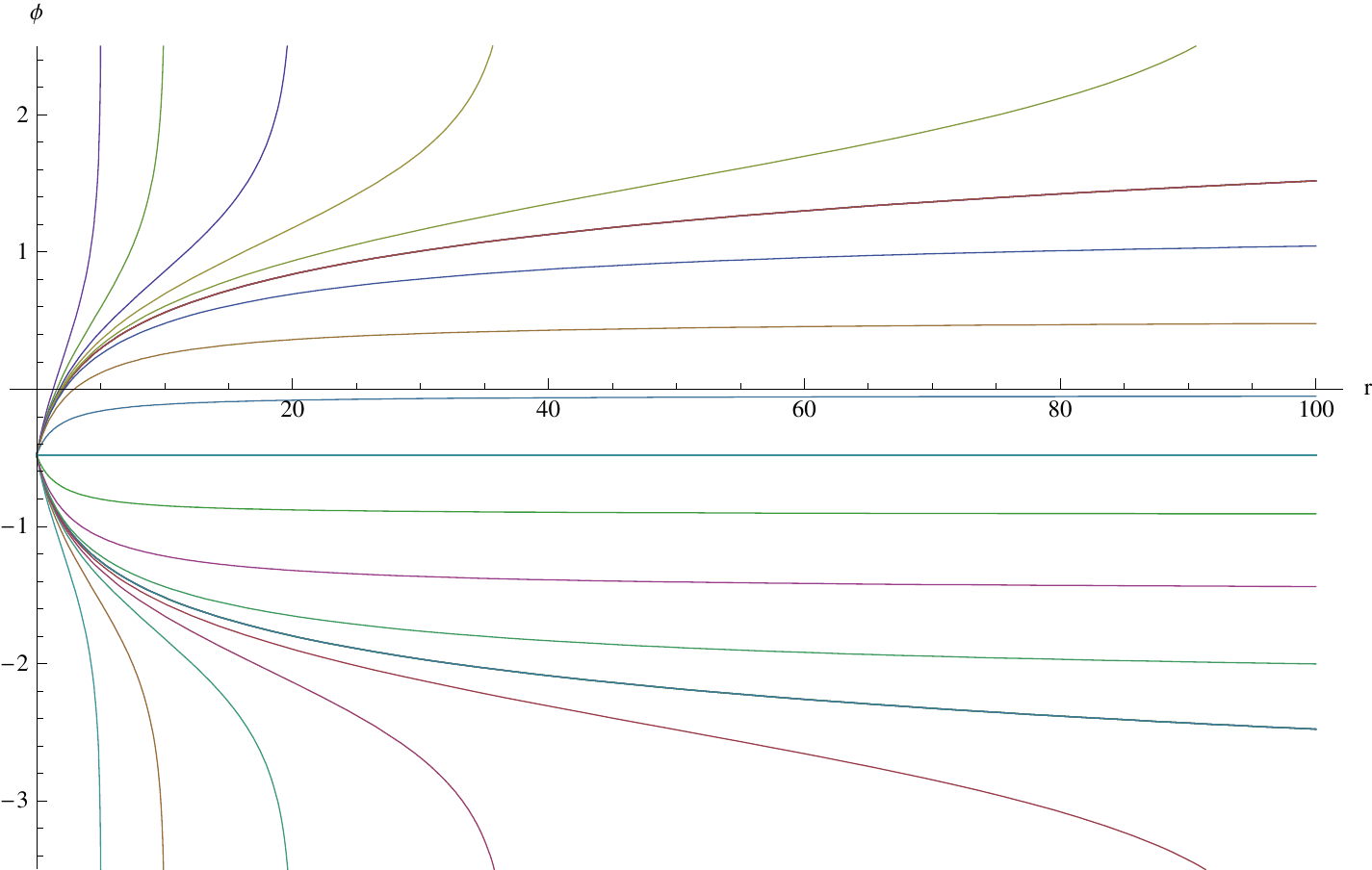}
\caption{The attraction basin for our toy model when $Q_1=7, Q_2=4$. The attractor value of the scalar is $\frac{\sqrt{3}}{4}\log\Big(\frac{Q^2_m}{\sqrt{3}\ Q^1_m}\Big)=-0.480177$. The upper region where the curves diverge at finite radius correspond to $d_1 <0, d_2>0 $, the lower region where they diverge is where $d_2 <0, d_1>0$. Inside the attraction basin, it is $d_1, d_2 >0$ and the upper (lower) boundary is $d_1=0$ ($d_2=0$). The last non-divergent curves are logarithmic in $r$ as $r \rightarrow \infty$, and correspond to generalized subttractors. The ``hairless" black hole corresponds to $d_1=d_2$. We have chosen to make this plot with $d=\sqrt{d_1^2+d_2^2}=1$, and then tuning $\theta$ defined by $d_1=d \cos \theta,  \ d_2 = d \sin \theta$. The different curves in the figure correspond to different values of $\theta$. The qualitative features of the plot will not change for other numerical values of $d$ and more generally, as long as we are using a monotonic curve in the $d_1$-$d_2$ plane to slice through the attraction basin.  
}
\end{center}
\end{figure}

We would like to scan the solutions for various asymptotic values to get an intuition about the attraction basin. The above equations suggest that this is equivalent to scanning (non-negative) $d_1$ and $d_2$. However, since the axes of the 1st quadrant in the $d_1$-$d_2$ plane corresponds to the boundaries of attraction, there is a whole (non-negative) $d_1$ ray worth of solutions corresponding to the $d_2=0$ boundary (and vice-versa). So to slice the attraction basin and see both boundaries in an instructive plot, it is crucial that we move in the $d_1$-$d_2$ plane along a curve that cuts both (positive) axes\footnote{Note also that the hairless black hole corresponds to the $d_1=d_2$ solution, the central positive ray in the $d_1$-$d_2$ plane.}. In particular, one will not be able to see the other boundary by starting with the $d_2=0$ boundary for some value of $d_1$ and then cranking up $d_2$ arbitrarily (while holding $d_1$ fixed). To make the plot we present, we have chosen 
\bea
d_1=d \cos \theta,  \ \ d_2 = d \sin \theta,
\eea
with $d$ fixed, and then plotted the scalar $\phi$ as a function of $r$ for various values of $\theta$. 

The above choice of the curve was arbitrary. But starting with a given black hole solution fixes the choice of $d_1$ and $d_2$ (one might call these $d_1^0$ and $d_2^0$), and perturbations around this solution are then determined by the leading order scalar perturbation, which we will call $\phi_1$. Since a general solution is fixed once we know $d_1$ and $d_2$, we can determine the $d_1$ and $d_2$ of the perturbed solution in terms of $\phi_1, d_1^0$ and $d_2^0$ by matching it order-by-order with  the near-horizon series expansion of the general solution (\ref{singlescalar}-\ref{singlescalar2}). Since $d_1$ and $d_2$ are determined in terms of one parameter $\phi_1$, this picks out a unique curve through the attraction basin with initial values $d_1^0$ and $d_2^0$. We explain this in detail in Appendix \ref{subtslice} and demonstrate with the example where our starting solution is the subttractor geometry of \cite{Avik}, given by a specific choice of $d_1^0$ and $d_2^0$.

We discuss some of the features of the plot in the caption of figure 1, and similar statements hold also for the more general STU system we considered in the previous sections. The structure of the attraction basin is such that as one of the $d_a$'s goes through zero and becomes negative, the solutions no longer have an asymptotic region where $r \rightarrow \infty$. This is because (as we mentioned) the $(1+d_a r)$ causes a divergence at $r=-1/d_a\ (>0) $ when $d_a$ is negative, and the solutions literally blow up at finite radius. Note however that the extremal near horizon geometry and the attractor value of the scalar are still preserved - so the near-horizon region is {\em still} attractive to the scalars even though they are blowing up and losing their asymptotic region\footnote{Since crossing over the $d_a=0$ boundary brings out this conflicting, self-destructive behavior from scalars, one might call them ``dramatic 
solutions". But perhaps one should not judge these solutions {\em too} harshly.}.  

\section{Comments}
\label{comments}

We will conclude with some comments, open questions, future directions, etc. Some of these questions are being investigated \cite{new}.

We have presented generalized subtracted geometries in the extremal limit. These have a straightforward generalization to the non-extremal case, even though we have not emphasized this in our paper. However, both these results are limited to the case of static black holes. What would be more interesting is if the generalized subtracted geometry also exists in the case of black holes with rotation. In \cite{CL2}, the warp factor for spinning black holes was chosen by demanding that the wave equation is separable. Together with the demand for asymptotic linear behavior, this was enough to essentially uniquely fix the warp factor of the subtracted geometry. For higher powers in the asymptotic behavior, it seems likely that the separability criterion is not enough to uniquely fix the warp factor\footnote{See \cite{Larsennew} for a discussion of separability of black holes in string theory.}. However, it will be interesting to see how constraining separability is. Together with our knowledge of the warp factor in the static limit, and by demanding that the equations of motion be satisfied, it might be possible to determine a separable, generalized subtracted geometry for rotating black holes as well. A related and very likely possibility is that one might be able to generate these new solutions via an appropriate Harrison transformation \cite{CG, Amitabh}\footnote{See \cite{Klemm} for an early use of the Harrison transformation.}. Finally, there is the question of constructing attraction boundaries for rotating extremal black holes \cite{Kevin} which is a natural extension of these ideas. 

The solutions we have investigated all were integrable via the Toda trick. Furthermore, the toy model we considered is secretly connected to the STU model \cite{Avik}, so it is not necessarily an independent check of the basin structure.  Our goal in presenting it was mostly for visualization and intuition, since the STU model with its many scalars and charges and parameters is rather unweildy. We have not investigated the question of robustness of the basin structure in other systems. These include other systems that allow Toda-like solutions (cf. \cite{Goldstein}'s appendices) as well as the large class of Einstein-Maxwell-Dilaton theories that one can cook up\footnote{Even though many of them might not be interesting from the point of a top-down string theory/supergravity picture.} that are not necessarily integrable in this simple way. 

Recently, there has been some discussion where the attractor mechanism has been connected to gauged supergravity and also to the possibility of anisotropic geometries \cite{sandip1, sandip2}. It will be interesting to investigate similar questions as we have done here, in these contexts.

A comment worth making is the connection with supersymmetric attractors. It seems plausible that the discussion here can be presented in a form that has analogies with the supersymmetric discussion. The solutions here have some moral similarities with the susy discussion \cite{Kallosh, Sabra}. 

A question that we have not addressed is that of stability. To the extent that non-supersymmetric attractors are stable, we believe the discussion in this paper is safe. But it will of course be interesting to investigate the stability of these solutions in the full theory. 
It is worthwhile noting here that what we have shown here and in \cite{Avik} amounts to a certain kind of perturbative instability of the (extremal) subtracted geometry. This is indeed what it means to say that the subttractor is a boundary between two classes of solutions. 

\section*{Acknowledgments}

We would like to thank Sanjib Jana and Avinash Raju for discussions. CK thanks the organizers and participants of ISM 2012, Puri, for an inspiring conference.


\appendix

\section{General Subttractors}
\label{gsubtt}

Here we present the explicit forms of the metrics and scalars on the various attraction boundaries and make some comments about them. To avoid repeating long expressions too much, we first define
\bea
H_a(r)=Q_a (1+d_a \ r).
\eea
Note that this is a linear polynomial in $r$, our solutions will essentially be made out of powers of them. In what follows, all non-zero $d_a$'s are assumed positive. For comparison with other papers, we define the warp factor $\Delta$, as $b^2(r)=\Delta(r)^{1/2}$.  
\begin{itemize}
\item[$\diamondsuit$] $d_4 \neq 0$ case.
\begin{itemize}
\item[$\bullet$] $d_1=d_2=d_3=0$.
\bea
e^{4 \phi_1}=\frac{Q_1\ Q_3}{Q_2 \ H_4},\ 
e^{4 \phi_2}=\frac{Q_2\ Q_3}{Q_1 \ H_4},\ 
e^{4 \phi_3}=\frac{Q_1\ Q_2}{Q_3 \ H_4}, \ a^2=\frac{1}{4}\frac{(Q_1Q_2Q_3)^{-1/2}\ r^2}{\sqrt{H_4}}.\label{standardsubtttractor}
\eea
In the language of \cite{CL2} this corresponds to working with an asymptotically linear warp factor in the metric (ie., $\Delta \sim r$) and is the standard (extremal) subtracted geometry \cite{Avik}.   
\item[$\bullet$] $d_1=d_2=0$, $d_3 \neq 0$.
\bea
e^{4 \phi_1}=\frac{Q_1\ H_3}{Q_2 \ H_4}, \
e^{4 \phi_2}=\frac{Q_2\ H_3}{Q_1 \ H_4}, \
e^{4 \phi_3}=\frac{Q_1\ Q_2}{H_3 \ H_4}, \
a^2=\frac{1}{4}\frac{(Q_1Q_2)^{-1/2}\ r^2}{\sqrt{H_3 H_4}}.
\eea
The warp factor here goes as $\sim r^2$ in the asymptotic region.
\item[$\bullet$] $d_1=0$, $d_2 \neq 0 \neq d_3$.
\bea
e^{4 \phi_1}=\frac{Q_1\ H_3}{H_2 \ H_4}, \
e^{4 \phi_2}=\frac{H_2\ H_3}{Q_1 \ H_4}, \
e^{4 \phi_3}=\frac{Q_1\ H_2}{H_3 \ H_4}, \
a^2=\frac{1}{4}\frac{(Q_1)^{-1/2}\ r^2}{\sqrt{H_2 H_3 H_4}}.
\eea
The warp factor here goes as $\sim r^3$ in the asymptotic region.
\item[$\bullet$] None of $d_1, d_2, d_3$ are zero. This is the flat space extremal solution (\ref{extremalexact1}-\ref{extremalexact2}), not a boundary. For generic positive values of $d_a$ these are hairy flat space black holes. The warp factor goes as $\sim r^4$.
\end{itemize}
\item[$\diamondsuit$] $d_4 = 0$ case.
\begin{itemize}
\item[$\bullet$] $d_1=d_2=d_3=0$. This ``vertex of attraction" solution is covered in Appendix \ref{ads}. The warp factor is a constant, $\sim r^0$.
\item[$\bullet$] $d_1=d_2=0$, $d_3 \neq 0$.
\bea
e^{4 \phi_1}=\frac{Q_1\ H_3}{Q_2 \ Q_4}, \
e^{4 \phi_2}=\frac{Q_2\ H_3}{Q_1 \ Q_4}, \
e^{4 \phi_3}=\frac{Q_1\ Q_2}{H_3 \ Q_4}, \
a^2=\frac{1}{4}\frac{(Q_1Q_2Q_4)^{-1/2}\ r^2}{\sqrt{H_3}}.
\eea
This solution also has a warp factor that goes as $\sim r$ like the standard subtracted geometry, but the scalars that support it are not of the form adopted in \cite{CL2, CG}. 
\item[$\bullet$] $d_1=0$, $d_2 \neq 0 \neq d_3$.
\bea
e^{4 \phi_1}=\frac{Q_1\ H_3}{H_2 \ Q_4}, \
e^{4 \phi_2}=\frac{H_2\ H_3}{Q_1 \ Q_4}, \
e^{4 \phi_3}=\frac{Q_1\ H_2}{H_3 \ Q_4}, \
a^2=\frac{1}{4}\frac{(Q_1Q_4)^{-1/2}\ r^2}{\sqrt{H_2 H_3}}.
\eea
The warp factor here goes as $\sim r^2$.
\item[$\bullet$] None of $d_1, d_2, d_3$ are zero.
\bea
e^{4 \phi_1}=\frac{H_1\ H_3}{H_2 \ Q_4}, \
e^{4 \phi_2}=\frac{H_2\ H_3}{H_1 \ Q_4}, \
e^{4 \phi_3}=\frac{H_1\ H_2}{H_3 \ Q_4}, \
a^2=\frac{1}{4}\frac{(Q_4)^{-1/2}\ r^2}{\sqrt{H_1H_2 H_3}}.
\eea
The warp factor of this subttractor geometry goes as $\sim r^3$.
\end{itemize}
\end{itemize}

The asymptotic structure of all of these generalized subttractor metrics fall into the general form
\bea
ds_n^2=-\frac{r^{2-n}}{\sigma_n}dt^2+\frac{\sigma_n}{r^{2-n}}dr^2 +\sigma_n \ r^n d\Omega_2,
\eea
where $\sigma_n$ is a constant.
It is easy to translate this into a more conventional radial coordinate $R^2= \sigma_n \ r^n$, and we find
\bea
ds_n^2=-\frac{R^{\frac{4}{n}-2}}{\sigma_n^{\frac{2}{n}}}dt^2+\frac{4}{n^2}dR^2 + R^2 d\Omega_2
\eea
This is a conical box metric. The cases we care about here are $2n=1,2,3,4$. The case $2n=4$ corresponds to flat Minkowski space, $2n=1$ is the conical box of \cite{CG}. It is easily checked that the geometry has a curvature singularity at $R=0$ for all $2n \neq 4$.

\section{The $AdS_2 \times S^2$ solution}
\label{ads}

When all the $d_a=0$ in the STU solution we considered, the solution takes the form
\bea
e^{4 \phi_1}=\frac{Q_1\ Q_3}{Q_2 \ Q_4},\ 
e^{4 \phi_2}=\frac{Q_2\ Q_3}{Q_1 \ Q_4},\ 
e^{4 \phi_3}=\frac{Q_1\ Q_2}{Q_3 \ Q_4}, \\ a^2=\frac{r^2}{4\sqrt{Q_1Q_2Q_3Q_4}}\equiv \frac{r^2}{R_0^2}. \hspace{0.8in}
\eea
The explicit form of the metric is then
\bea
ds^2=-\frac{r^2}{R_0^2}dt^2+\frac{R_0^2}{r^2}dr^2+R_0^2 d \Omega^2
\eea
which is precisely of the $AdS_2 \times S^2$ form. This is the Freund-Rubin type reduction of our original theory where the sphere is supported by the gauge fluxes. 

This solution is therefore qualitatively different from the case where all $d_a$ are equal to each other but different from zero (and positive). Those solutions are asymptotically flat. It is natural to call this solution the vertex of attraction because in the non-negative $d_1$-$d_2$-$d_3$-$d_4$ orthant, this solution corresponds to the origin. 

\section{4D, Static, Extremal, String Theory Black Holes}
\label{4d}

The general (up to hair) four-charge static black hole in ${\cal N}=4$ string theory in 4 dimensions was constructed in \cite{CveticYoum2, Chong}. We will follow the conventions of \cite{CG}. This solution, in a form that is convenient for us is presented in eqns. (A.2-A.7) in \cite{Avik}. Our charges $Q_a$ are related to the parametrization used in \cite{CG, Avik} via
\bea
Q_a=\frac{m\sinh 2 \delta_a}{2}
\eea
which is slightly different from the normalization for the charges used in \cite{Avik}. The formula for the mass is
\bea
M=\frac{m}{4} \sum_a \cosh 2 \delta_a.
\eea
We will be interested in the extremal limit, where $m \rightarrow 0, \delta_a \rightarrow \infty$ while the charges $Q_a$ stay finite. In terms of the attractor ansatz, this solution is of the form (\ref{extremalexact1}-\ref{extremalexact2}) with $d_a= \frac{1}{2|Q_a|}$. The gauge field strengths take the form 
\bea
F_1=2 Q_1 \sin \theta d\theta \wedge d\phi, \ && \ F_2=\frac{2Q_2}{(r+2 Q_2)^2}dt \wedge dr, \\
F_3\equiv{\cal F}^{1}=2 Q_3 \sin \theta d\theta \wedge d\phi, \ && \ F_4\equiv{\cal F}^2=\frac{2Q_4}{(r+2 Q_4)^2}dt \wedge dr.
\eea

\section{Slicing Through the Attractor Basin}
\label{subtslice}

In this Appendix, we will do a near-horizon perturbation theory of the general solution in the single scalar toy model (\ref{singlescalar}-\ref{singlescalar2}). The solution that we perturb around will be described by some (non-negative) parameters $d_1=d_1^0$ and $d_2=d_2^0$, and the perturbations that are regular at the horizon are described completely by a single ${\cal O}(r)$ scalar perturbation parameter $\phi_1$. But since the perturbed solution should also be of the form (\ref{singlescalar}-\ref{singlescalar2}), the $d_1$ and $d_2$ of this perturbed solution should be expressible in terms of $d_1^0, d_2^0$ and $\phi_1$ (See eqns. (\ref{p1}-\ref{p2})). We can determine this relationship by doing a near-horizon expansion of the general $d_1$-$d_2$ solution and matching it with the perturbed (by $\phi_1$) near-horizon equations around the $d_1^0, d_2^0$ solution. 

In effect, $d_1^0, d_2^0$ act as initial conditions for the solution flow in the $d_1$-$d_2$ moduli space, and the flow curve is parametrized by $\phi_1$. Our goal in this section is to determine this curve. We will conclude by presenting the full plot of the solution flow along the attraction basin corresponding to perturbations around the subttractor solution considered in \cite{Avik}. This latter solution is a specific choice of $d_1^0, d_2^0$, and in figure 2 that we present, the various curves are various points in the attraction basin through which the perturbation from the $(d_1^0, d_2^0)$ solution moves. 

By demanding that they satisfy the equations of motion, the perturbed fields that are regular at the horizon can be expanded in a series as
\bea
a(r)=\alpha\Big[r - r^2\Big(\frac{d_1^0 + 3\ d_2^0}{
 4}\Big) + 
 r^3 \Big(\frac{ 5\ d_1^0{}^2 + 6\ d_1^0 \ d_2^0 + 21\ d_2^0{}^2}{32} + \nonumber\hspace{1in}\\ \hspace{0.3in}-\frac{
     \sqrt{3}\ d_1^0 \ \phi_1 -\sqrt{3}\ d_2^0 \ \phi_1 - 
       2 \ \phi_1^2}{4}\Big)+\dots\Big] \label{p1}\\
b(r)=\frac{1}{\alpha}\Big[1 + r\Big(\frac{d_1^0 + 3\ d_2^0}{
 4}\Big)+ \frac{r^2}{32} \Big(-3 \ d_1^0{}^2 + 6\ d_1^0\ d_2^0 - 3\ d_2^0{}^2+ \nonumber\hspace{1in}\\+ 8 \sqrt{3}\ d_1^0\ \phi_1 - 
   8 \sqrt{3}\ d_2^0 \ \phi_1 - 16 \ \phi_1^2\Big) +\dots \Big] \\
\phi(r)= \frac{\sqrt{3}}{4}\log\Big(\frac{Q_2}{\sqrt{3} Q_1}\Big)+r \Big(\frac{-\sqrt{3}\ d_1^0 + \sqrt{3}\  d_2^0}{4}+\phi_1\Big) + \nonumber\hspace{1.3in}\\+ \frac{r^2}{24} \Big(3 \sqrt{3}\ d_1^0{}^2 - 3 \sqrt{3} \ d_2^0{}^2 - 18\ d_1^0 \ \phi_1 - 
   6 \ d_2^0 \ \phi_1 + 8 \sqrt{3}\ \phi_1^2\Big)+\dots \label{p2}
\eea
where
\bea
\alpha\equiv\frac{3^{3/8}}{2 (Q_1 Q_2^3)^{1/4}}.
\eea
If one chooses to work with numerical solutions and start integrating these from the horizon, it is useful to have more subleading terms in these expansions for those values of the perturbation $\phi_1$ for which the ${\cal O}(r)$ term in the scalar expansion vanishes. It is straightforward to determine them by demanding that the equations of motion are satisfied order by order in $r$ and we have computed them, but they are too cumbersome and we will not present them. In any event, since we have the full analytic solutions, we can always translate any question we want in this context into an analytically tractable question.

\begin{figure}
\begin{center}
\includegraphics[height=0.35\textheight]{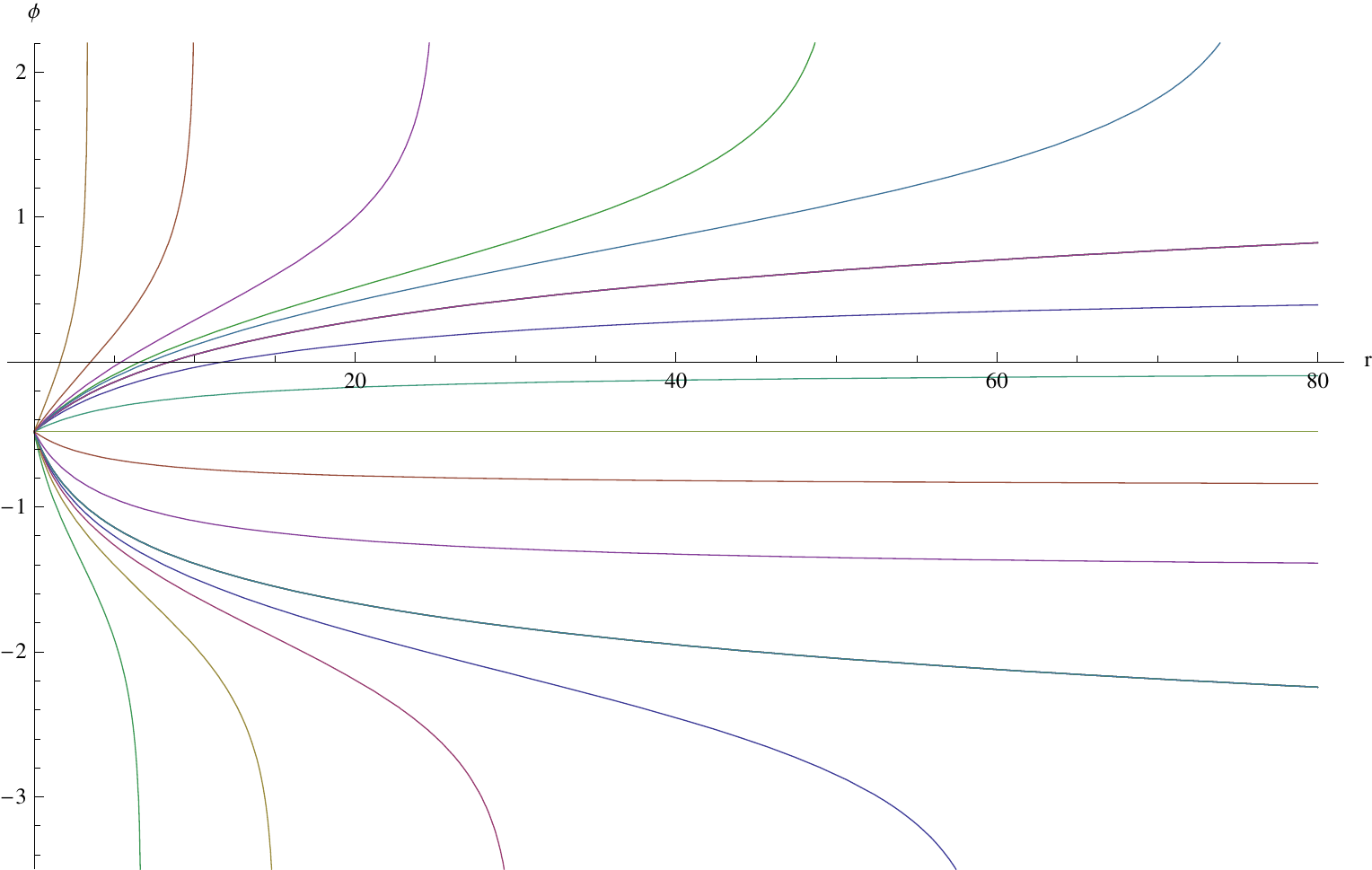}
\caption{The $\phi$-$r$ curves corresponding to the slice through the attraction basin for our toy model when $Q_1=7, Q_2=4$. This is the slice in which the subttractor presented in \cite{Avik} lies. As we have emphasized, the qualitative features of the plot are not different from our rather ad-hoc circular slicing of the basin presented in the main body of the paper.
}
\end{center}
\end{figure}

The near-horizon expansion of a general solution with parameters $d_1, d_2$ is
\bea
a(r)=\alpha\Big[r - r^2\Big(\frac{d_1 + 3\ d_2}{
 4}\Big) + 
 r^3 \Big(\frac{ 5\ d_1{}^2 + 6\ d_1 \ d_2 + 21\ d_2{}^2}{32}\Big)+\dots\Big] \hspace{0.3in}\\
b(r)=\frac{1}{\alpha}\Big[1 + r\Big(\frac{d_1 + 3\ d_2}{
 4}\Big)+ \frac{r^2}{32} \Big(-3 \ d_1{}^2 + 6\ d_1\ d_2 - 3\ d_2{}^2\Big) +\dots \Big] \hspace{0.3in}\\
\phi(r)= \frac{\sqrt{3}}{4}\log\Big(\frac{Q_2}{\sqrt{3} Q_1}\Big)+r \Big(\frac{-\sqrt{3}\ d_1+ \sqrt{3}\  d_2}{4}\Big) + \frac{r^2}{24} \Big(3 \sqrt{3}\ d_1{}^2 - 3 \sqrt{3} \ d_2{}^2\Big)+\dots,
\eea
and we want to match this with the previous perturbation expansion to determine $d_1$ and $d_2$ in terms of $d_1^0, d_2^0$ and $\phi_1$. It is easy to see that the entire series expansion matches term by term when we set
\bea
d_1=d_1^0-\sqrt{3} \phi_1, \ \ d_2=d_2^0+\frac{\phi_1}{\sqrt{3}} \label{look}
\eea
Eliminating $\phi_1$ (which is essentially the parametrization of the curve in the attractor basin), we get the slicing curve (in fact a line) through the $d_1$-$d_2$ attraction basin to which a black hole with parameters $d_1^0$ and $d_2^0$ belongs to:
\bea
d_2=\frac{1}{3}(d_1^0-d_1)+d_2^0 \label{sliceline}
\eea

As final illustration, we plot the flow in the attraction basin (in $\phi$-$r$ space) starting with the subttractor solution we considered in \cite{Avik}, which corresponds to
\bea
d_1^0=\frac{1}{4}\Big(\frac{6 \sqrt{3}}{Q^2}+\frac{2}{Q^1}\Big), \ \ \ d_2^0=0.
\eea
We can do this by numerically integrating the perturbations carefully as we mentioned before, or by plotting the various solutions along the curve (\ref{sliceline}) with $d_1^0$ and $d_2^0$ chosen as above. The result is plotted below and as we emphasized, the qualitative features are identical to what we saw in the main body of the paper. 

Similar constructions can be done for the 3-scalar solutions in the STU model as well. We won't present the details of the perturbation theory because it is merely more complicated and not conceptually different. But it turns out that the hyperplane that cuts through the attraction basin in $d_1$-$d_2$-$d_3$-$d_4$ space under perturbations of a black hole specified by $d_1^0$, $d_2^0$, $d_3^0$, $d_4^0$ is given by the simple relation
\bea
d_1+d_2+d_3+d_4=d_1^0+d_2^0+d_3^0+d_4^0.
\eea
Note that in the STU model there are three independent scalar perturbation $\phi_1, \phi_2, \phi_3$ now, which correspond to different possible directions on the hyperplane for the solutions to move. To obtain the equation above, we have eliminated these perturbations from 
\bea
d_1=d_1^0 + \phi_1 - \phi_2 + \phi_3, \ \  d_2 = d_2^0 - \phi_1 + \phi_2 + \phi_3, \\ d_3 = d_3^0 + \phi_1 + \phi_2 - \phi_3, \ \ d_4 = 
 d_4^0 - \phi_1 - \phi_2 - \phi_3.
\eea
These are the analogues (\ref{look}) in the STU system.


%

\end{document}